\documentstyle[draft,epsf]{mn}

\def\hii{{\rm H{\sc ii} }}
\def\gsim{\mathrel{\raise0.35ex\hbox{$\scriptstyle >$}\kern-0.6em 
\lower0.40ex\hbox{{$\scriptstyle \sim$}}}}
\def\lsim{\mathrel{\raise0.35ex\hbox{$\scriptstyle <$}\kern-0.6em 
\lower0.40ex\hbox{{$\scriptstyle \sim$}}}}

\title{The Colour-Magnitude Relation of Early-Type Galaxies in
        the Hubble Deep Field}
\author[Kodama, Bower \& Bell]
{
Tadayuki Kodama$^1$, Richard G. Bower$^2$ \& Eric F. Bell$^2$ \\
$^1$ Institute of Astronomy, University of Cambridge, Madingley Road, Cambridge CB3 0HA, UK\\
$^2$ Department of Physics, University of Durham, South Road, Durham DH1 3LE, UK}

\begin{document}

\maketitle

\begin{abstract}
We present the rest-frame colour-magnitude diagram for 35 early-type
galaxies in the Hubble Deep Field with median redshift 0.9.
Although with considerable scatter, a red sequence well-described by
the passive evolution of an intrinsically old stellar population is
observed.  Comparison with the passively evolved colour-magnitude
relation of the rich Coma cluster ($z=0.023$) indicates that around
half of the early-type 
galaxies in the field at this redshift are as old as those
in rich clusters.
\end{abstract}

\begin{keywords}
galaxies: elliptical -- galaxies: evolution -- galaxies: stellar content
\end{keywords}

\section{Introduction}

Early-type galaxies in clusters exhibit a strong correlation between their
absolute magnitudes and colours (eg., Visvanathan \& Sandage 1977;
Bower, Lucey, \& Ellis 1992, hereafter BLE92). This correspondence is referred
to as the colour-magnitude relation (CMR).
The tightness of the CMR can be used to set important 
constraints on the star formation history of these galaxies, indicating
an impressive degree of homogeneity in the formation process.
The correlation is most
naturally accounted for as a sequence of metal abundance, with the scatter
being primarily driven by differences in galaxy age. This paradigm
correctly predicts the evolution of the zero-point and slope of the
relation with redshift (Kodama et al. 1998a; Stanford et al. 1998), and
is supported by line index analysis of galaxies in the Fornax and Coma 
clusters (Kuntschner \& Davies 1997; Terlevich 1998).
Bower, Kodama \& Terlevich (1998; hereafter BKT98) have recently reviewed the
limits on formation history that can be derived from local clusters.
They conclude that the small
scatter of the relation implies that the bulk of the stellar populations
are old, most of the stars having been formed before a look-back time
of 9~Gyr ($z\sim1$ with $H_0=50$~km/s/Mpc and $q_0=0.5$);
although some residual star formation in a fraction
of the galaxies is {\it required} to account for the increasing fraction
of blue galaxies observed in the core of intermediate redshift clusters
(eg., Butcher \& Oemler 1984)

A key question is to determine whether the same homogeneity that is seen 
in the clusters extends to galaxies in lower density environments. 
The available evidence indicates that the CMR is present in early-type 
galaxies in compact groups (Zepf et al. 1991). However, comparable 
data-sets for genuine `field' E and S0 galaxies
(ie., galaxies selected along an average line of sight) is sparse.
One of the most homogeneous studies is that of Larson, Tinsley \&
Caldwell (1980). They found that scatter of field Elliptical galaxies was
greatly enhanced over equivalent galaxies in clusters; however, the
same trend was not seen in the S0 galaxy population. More recent
studies have concentrated on spectroscopic indicators of stellar
populations. For instance, Guzman et al. (1992) report a small offset
in the normalisation of the Mg$_2$-$\sigma$ correlation of early-type
galaxies in the core and outskirts of the Coma cluster. However,
the implied differences in mean stellar population ages are smaller than
10 per cent.

In this paper we address the homogeneity of the formation
of early-type galaxies in low density environments using the colours of 
morphologically classified E
and S0 galaxies in the Hubble Deep Field (HDF; Williams et al. 1996).
The galaxies are typically at
redshifts of order unity, allowing us to take advantage of the large
look-back time.  Despite the large distances to these objects,
the narrow point spread function of the Hubble Space Telescope allows
accurate morphological classification. 
Furthermore, because these galaxies 
cover only a small area of sky, uncertainties in the photometric 
zero-point and galactic extinction do not introduce spurious scatter into 
the relation as they might do in the local surveys.

The plan of the paper is as follows. In \S~2, we describe the method
used to create the rest-frame CMR. The star formation
history is derived by comparing the colour distribution with the model
in \S~3 and the discussion is given in \S~4.  Conclusions are given in \S~5.

\section{The Field Colour-Magnitude Relation at $z=0.9$}

\subsection{Early-type galaxies in the Hubble Deep Field}

Franceschini et al. (1998; hereafter F98) define a sample of
early-type galaxies in the HDF. We will regard this sample as
representative of the field early-type galaxy population  
at  $z\sim0.9$. Their sample contains 35 morphologically classified 
early-type galaxies. Their initial selection is based on $K$-band
surface brightness profiles from the KPNO-IRIM images of
Dickinson et al.\ (1997), with further checks on the surface brightness
profile being made using HST $V$ or $I$ band data (Fasano et al. 1998).
The sample galaxies are flux-limited with $K<20.15$.
F98 present photometry in  7 passbands from the UV to NIR 
($U_{300}$,$B_{450}$,$V_{606}$,$I_{814}$,$J$,$H$,$K$) using a fixed 
$2.5\arcsec$ diameter aperture. 

\subsection{Rest-frame colours and magnitudes}

Fifteen out of the 35 F98 galaxies have spectroscopic redshifts
available from the literature (Cohen et al. 1996; Cowie 1998). 
For the remainder redshifts must be estimated using
photometric redshift techniques. We use the estimator of Kodama, Bell \&
Bower (1998b). At $z \sim 1$, the bluer optical colours primarily
reflect the colours of the rest frame far UV, where model colours are
highly uncertain, and we therefore adopt an iterative redshift
estimation scheme. 
We used all seven passbands from $U_{300}$ to $K$ for the first estimation.
For the second estimation, we excluded the passbands which would fall
below rest frame 2500~\AA\, assuming that the first galaxy redshift
estimate is reasonably close to the final redshift estimate.
We used the 15 galaxies with spectroscopic redshifts in order to assess
the uncertainty of the technique. 
The agreement between our estimates and the spectroscopic ones are generally
good with redshift errors of $\overline{|\Delta z|}\sim0.08$. This
uncertainty is propagated through to the rest frame colours and
luminosities as described below.
As a final check, we confirmed that our estimates are also consistent
with F98's, again with $\overline{|\Delta z|}\sim0.1$.
So-called {\it catastrophic} redshift errors can occur
when the spectra of the model templates
are a poor match to the observed colours (Yee 1998; Kodama et al.\
1998b), which happens predominantly with star forming
late-type galaxy spectra.  In this analysis, we expect that
the redshift estimates of the redder, early-type galaxies will be
robust to this type of error, but we bear in mind that the
redshifts of the fainter, blue galaxies in our sample may be
susceptible to this source of error.
Our code also gives an estimate of bulge-to-total light (B/T) ratio
as well as the redshift for each galaxy (Kodama et al. 1998b), which we
later use to determine the K-correction for the galaxy.

K-corrections are applied to the data using evolutionary models for
early-type galaxies (in which chemical evolution is taken into account) 
from Kodama et al.\ (1998a; 1998b).  
The model galaxy has an old bulge component with mean stellar
metallicity $\langle\log Z/Z_{\odot}\rangle =-0.23$, and a disk
component.  The star formation rate in the disk is regulated by the 
Schmidt (1959) law with a star formation 
timescale of 5~Gyr. The formation redshift of the galaxies is
fixed at $z_{\rm form}=4.4$.
The relative contribution of the bulge and disk component in a given galaxy
(parameterised by the B/T ratio) is estimated along with the galaxy
redshift, and this is then used to estimate the K-correction.
In practice, the galaxies lying near the red envelope are unaffected by
the details of star formation in the model galaxy since they are
dominated by bulge light. Galaxies for which this correction process 
is potentially important lie far from the ridge-line of the main relation. 
Rest frame $U$-$V$ colours are estimated by K-correcting the observed
colours in passbands $(M_1,M_2)$ chosen to bracket the rest frame
4000~\AA\, break. We use
$(M_1,M_2)=(U_{300},V_{450})$ for $z < 0.2$,
$(B_{450},I_{814})$ for $0.2 \leq z < 0.6$,
$(V_{606},J)$ for $0.6 \leq z < 1.3$,
$(I_{814},J)$ for $1.3<z<1.6$,
and $(I_{814},H)$ for $z \geq 1.6$.

Because the sample is at a range of different redshifts in the field,
it is also important to apply evolution corrections (E-corrections),
so that a single coeval CMR can be constructed.  
For the sake of simplicity we assume, to first order, 
that all the galaxies are formed in a single burst at $z=z_{\rm form}$.
Using this model, we transform the colours of all the galaxies to the
median redshift of the sample to minimise the uncertainty in the E-correction.
However, it is clear from the range of galaxy colours that the evolution
of individual galaxies will not be well described by this model. 
For instance, the amount of E-correction for the bluer
galaxies at lower redshift ($z<0.9$) will be underestimated,
given that these galaxies are supposed to be younger.
The colour should evolve more rapidly, be bluer, and the colour deviation
from the Coma CMR at $z=0.9$ should be larger, compared to the
passive evolution of old stellar population.
We therefore
use the same passive evolution correction for all the galaxies
as a first approximation, and then apply an additional (smaller) correction
according to the colour deviation from the red sequence. 

We adopt a B/T$=$1.0 model and $z_{\rm form}=4.4$ for the primary
passive evolution correction.  The deviation of the galaxy from the 
Coma cluster CMR at $z=0.9$ (see below) is then 
used to estimate the secondary correction by multiplying by
a correction factor:
\begin{equation}
\Delta(U-V)_{\rm corrected}=\Delta(U-V)\times
\frac{\displaystyle t_{z=z_g}-t_{z=4.4}}{\displaystyle t_{z=0.9}-t_{z=4.4}},
\end{equation}
where $t_z$ indicates the time from the Big Bang at a given redshift.
This correction is adequate since the integrated colour $U-V$ of a galaxy
scales roughly logarithmically with age. For galaxies, lying within
$|\Delta (U-V)|<0.5$ of the fiducial CMR, this correction amounts to
only $\lsim 0.1$.
We apply this additional correction only to the colour.
As the slope of the CMR is quite shallow, the corresponding magnitude
correction has a negligible effect on the final results, and is
therefore neglected for the sake of simplicity.

The key aim of this work is to examine the spread of galaxies around
the CMR defined by rich clusters. Clearly, the photometric redshift
estimation method cannot determine a galaxy's redshift precisely. All
the galaxies analysed here have photometry of sufficient quality 
that the dominant source of uncertainty is the match between the object
and model spectrum. We therefore adopt the rms difference between the
spectroscopic and photometric determinations (0.08) as the uncertainty in the
photometric redshifts of the remaining part of the sample.
Although the spectroscopic redshifts are well determined, the rest-frame
colours are still uncertain because of the metallicity of the stellar
population.
The age of the stellar population also introduces
uncertainty, but this is degenerate with the uncertainty in metal
abundance (Worthey 1994, Kodama et al.\ 1998b).
To estimate these effects,
we applied the K+E-corrections using other elliptical galaxy
models with different metallicities in a range of
$-0.52 \leq \langle\log Z/Z_{\odot}\rangle \leq 0.06$,
but the resultant colours and magnitudes differ by less than 0.1 mag.
This shows that the corrections that we have made are relatively
insensitive to the metallicity variations that drive the CMR.

Because of the evolutionary corrections, our results depend slightly on
the adopted cosmological framework.
For our fiducial model, we adopt an open universe with 
($H_0,q_0$)$=$(64,0.1).
We also consider a flat universe with ($H_0,q_0$)=(50,0.5).
Both models give the same age of the universe at 13~Gyr. Hereafter, we
use the parameter $h$ to mean $H_0/100$.

\subsection{The colour-magnitude relation}

\begin{figure*}
\begin{center}
  \leavevmode
  \epsfxsize 0.7\hsize
  \epsffile{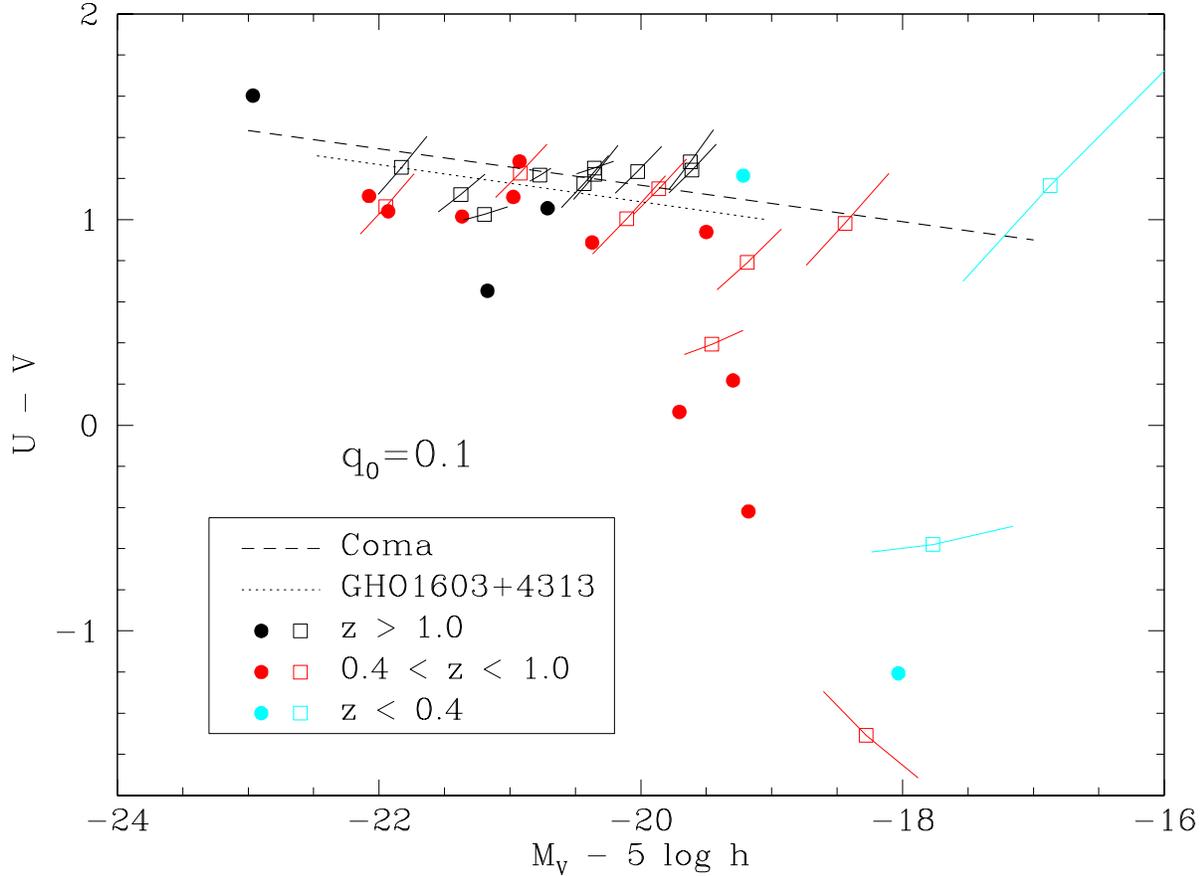}
\end{center}
\vspace{-3.5cm}
\caption{
The colour-magnitude diagrams for early-type galaxies
in the Hubble Deep Field observed at $z=0.9$ with a $q_0=0.1$ cosmology.
Filled symbols indicate galaxies with
spectroscopic redshifts, open symbols refer to those with 
photometric redshifts only.
The gray scale of the symbols indicates their redshift.
The dashed line shows the CMR of Coma transformed to $z=0.9$ using a passive
evolution correction.
The dot-dashed line shows the CMR of a cluster
GHO~1603+4313 at $z=0.9$ from Stanford et al. (1998) after K-correction.
}
\label{fig:cmr_q01}
\end{figure*}

The final $z = 0.9$ rest-frame colours and magnitudes are shown for 
the fiducial model in Fig.~\ref{fig:cmr_q01}. Galaxies with
spectroscopically determined redshifts are shown as points, where the
gray scale indicates the observed galaxy redshift.
Galaxies with photometric
redshifts are shown as error bars representing the 
estimated uncertainty
in the rest-frame colour and absolute magnitude.

{\it The ridge-line of the CMR is well defined even in field early-type
galaxies at $z=0.9$.}
This is a striking result, since the existence of a well
defined old stellar population at high redshifts seems contradictory
to the spirit --- if not the details --- of hierarchical galaxy
formation models (eg., White \&
Frenk 1991, Kauffmann, White \& Guiderdoni 1993, Cole et al.\ 1994,
Baugh et al.\ 1998, Kauffmann \& Charlot 1998a). 
An important test is the comparison of field and rich cluster CMRs.

Firstly, we compare the HDF CMR with the Coma cluster. The $z=0$ Coma cluster
relation is redder than the observed envelope by $\sim0.3$ mag at a
fixed luminosity. To make the comparison fair, we need to apply
a correction for passive evolution in order to determine the position
of the ridge-line of the Coma relation at $z=0.9$. This correction
results in the reddest possible CMR at high redshift.
To make the comparison, we use the $U-V$ data from BLE92.
However, because of the possibility of colour gradients, we must be 
careful to match the apertures in which colours are measured.
To match the aperture adopted by F98 we use $25\arcsec$ apertures 
(8.4~$h^{-1}$~kpc) for the $U-V$ colour and $32\arcsec$
(10.7~$h^{-1}$~kpc) apertures for the $M_V$ magnitude for the local Coma CMR.
$25\arcsec$ is the maximum aperture that allows us to reliably
measure the $U-V$ colour.
The large aperture Coma CMR is overplotted in Fig.~\ref{fig:cmr_q01}.
For the fiducial $q_0=0.1$ cosmology, the CMR just traces the reddest 
boundary of the HDF early-types. This suggests that at least some early-type
galaxies in the HDF must be as old as those in rich clusters (provided,
of course, that they obey  the same mass-metallicity relation).
We attempt to quantify this in \S3.
For $q_0=0.5$, the Coma relation lies blueward of many of the HDF
galaxies. Taken at face value, this implies that many of the field
galaxies formed their stars before the galaxies in rich clusters.

For further comparison, the CMR of a distant cluster GHO~1603+4313
at $z=0.9$ from Stanford et al. (1998) is also presented by applying
a K-correction using $(M_1,M_2)=(R,J)$ and eqs. (1) and (2)
in the same way as described in \S~3. In this case, aperture and  
E-corrections are not required. The location of the relation is again 
consistent with the red early-types
in the HDF, although there exists a slight systematic blue offset of order
$\sim$0.1 mag with respect to the Coma CMR. Although this offset
could result from observational zero-point mismatches and/or the slight model
uncertainties, if we adopt this relation rather than the transformed
Coma cluster relation, the HDF galaxies get older,
or more metal rich than their cluster counterparts.
We note that the offset disappears in the $q_0=0.5$ cosmology.

Although the overall impression is of a well defined CMR, there are
two important differences between this relation and that seen in rich 
clusters. Firstly, the relation shows little evidence of a slope (although
this may be masked by the large scatter). Any attempt to measure the
slope is dominated by the very bright, red galaxy 4-752-1. 
Because the slope is so poorly determined, we adopt the slope of
the Coma cluster CMR in what follows. Secondly,
the relation appears to have larger scatter.  This impression remains 
even if only the objects with spectroscopic redshifts are considered.
A maximum likelihood analysis of all the galaxies that have 
$U-V>0.7$  suggests an intrinsic scatter of 0.12$\pm$0.06 after allowing
for uncertainties in the individual data points.
This number, while uncertain, 
looks to be larger than the corresponding cluster value 
of $\sim$0.07~mag in the rest frame $U-V$ at the same redshift $z\sim0.9$
(Stanford et al. 1998).
Therefore the
scatter in the field, which is almost certainly {\it no less} than
the corresponding scatter for the $z = 0.9$ cluster, allows an epoch of
star formation similar to, if not more extended than, the epoch of
major star formation in cluster early-types.  This is in 
qualitative agreement
with hierarchical models. 

\section{Constraints on star formation history}

In order to quantify the constraints that are placed on the star
formation history of HDF early-type galaxies, we compare the
distribution of colour deviation from the $z = 0.9$ Coma CMR with the
predictions of a number of simple models.
As discussed in the introduction, we adopt a paradigm 
in which the colour magnitude relation in clusters is primarily driven
by universal correspondence between mass and metal abundance 
(Kodama \& Arimoto 1997; Kodama et al.\ 1998a; BKT98).
Following BKT98, we consider two types of star formation history: a
single burst model and  
a constant star formation (SF) model, truncated at
some epoch.
We randomly assign a single burst age
$t_{\rm burst}$ between a given set of $t_{\rm burst,min}$ and
$t_{\rm burst,max}$,
or a truncation time $t_{\rm stop}$ between $t_{\rm stop,min}$
and $t_{\rm start}$, respectively.
Note that $t_{\rm burst,max}$ and $t_{\rm start}$ should be less than the
age of the Universe at $z=0.9$, which is 6.0~Gyr for
($H_0$,$q_0$)$=$(64, 0.1) and 5.0~Gyr for (50, 0.5), respectively.
We fix the stellar metallicity at solar abundance,
since we compare colour residuals from the $z = 0.9$ CMR with the
models (therefore automatically taking out, to first order, the
metallicity dependence in this comparison).

Given the significant scatter of the dataset, the most appropriate
method of comparing the observed colour distribution with the
models is to create a cumulative histogram, and to apply
a Kolmogorov-Smirnov (K-S) test to the various model distributions.
This test requires that the models provide accurate {\it absolute}
colours, but this should not present a problem, as our model has been 
demonstrated to give the correct absolute colours for distant clusters
out to $z\sim1.2$ (Kodama et al. 1998a).
We compare the model colours with the observed ones using CMR residuals.
In the comparison, we convolve a Gaussian error of $\sigma = \pm 0.1$~mag
with the model colour in order to allow for uncertainties in the
K-corrected colours caused by redshift estimation uncertainties for the
red envelope galaxies. 

\begin{figure*}
\begin{center}
\centerline{
  \leavevmode
  \epsfxsize 0.5\hsize
  \epsffile{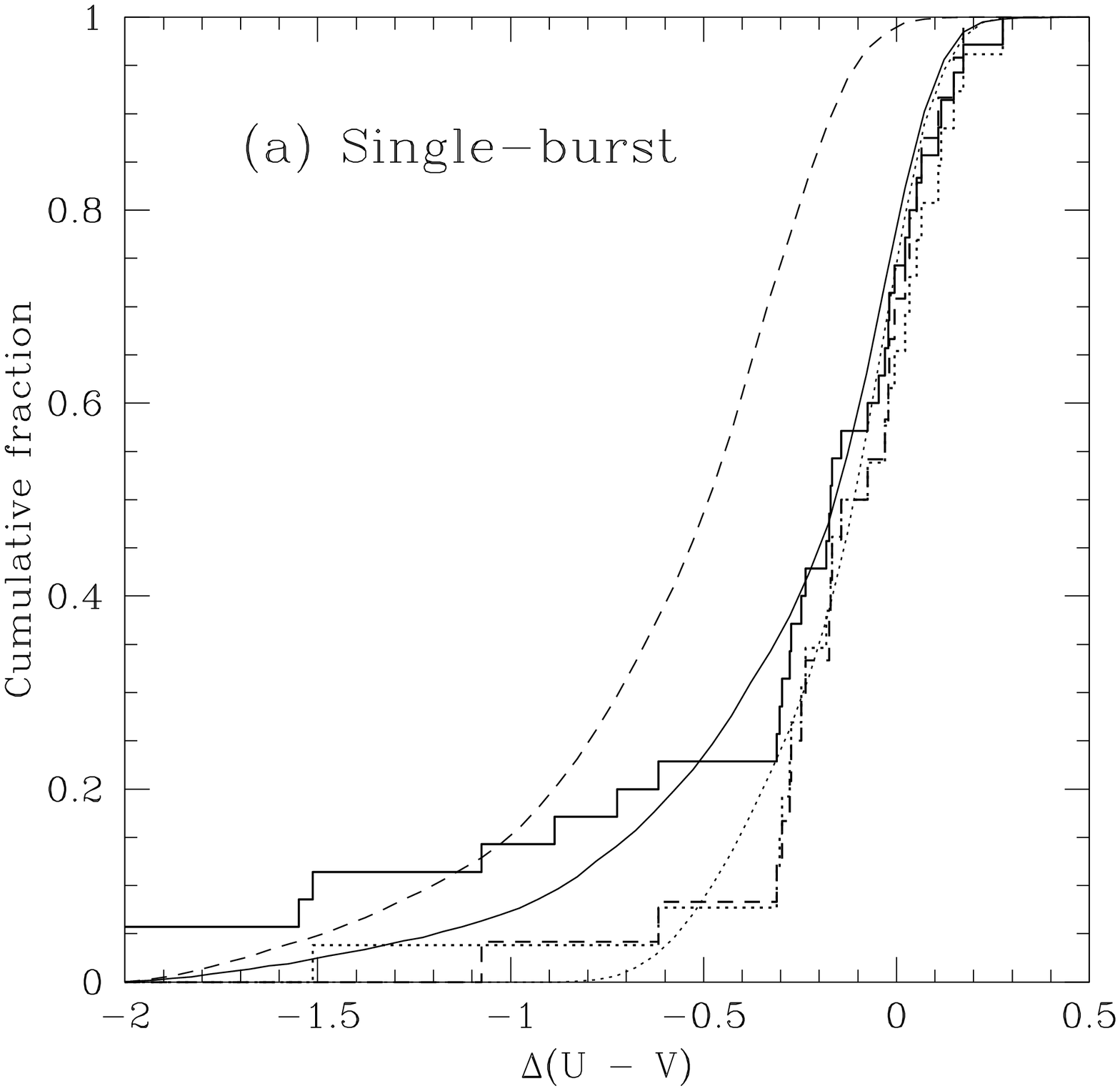}
  \leavevmode
  \epsfxsize 0.5\hsize
  \epsffile{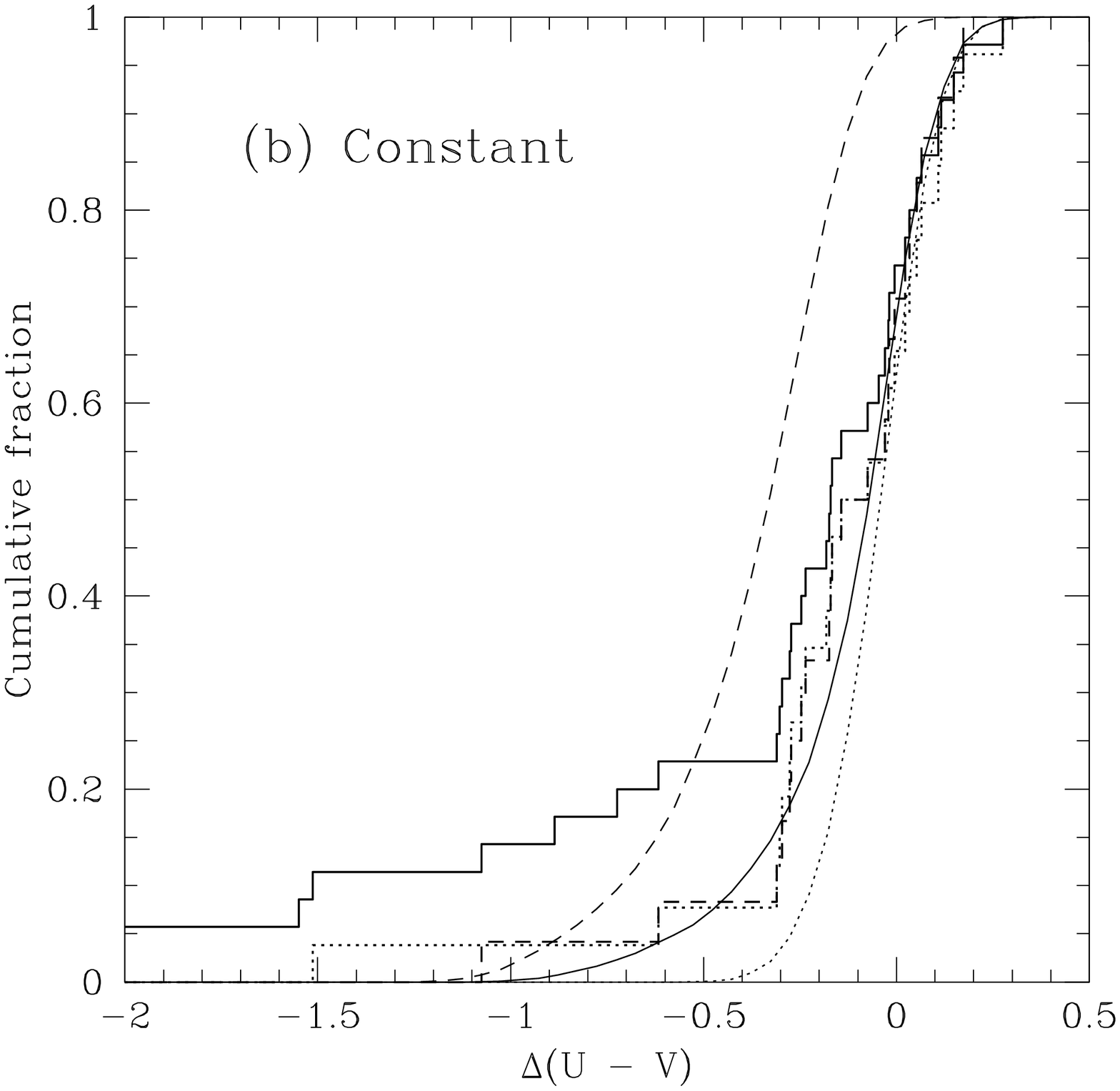}
}
\end{center}
\caption{
Cumulative histograms of the $U-V$ residuals from the K-corrected Coma CMR
($q_0=0.1$).
Observed histograms shows the all galaxies (solid histogram),
the bright galaxies only (dashed histogram), and the galaxies without
peculiars (dotted histogram).
Single burst models and the constant SF models are superposed
in the panel (a) and (b), respectively.
The models shown have ($t_{\rm burst,min}$ or $t_{\rm stop,min}$,
 $t_{\rm burst,max}$ or $t_{\rm start}$)=
(0, 6) [solid curve], (0, 3) [dashed curve], and (1, 6) [dotted curve].
}
\label{fig:cumhist}
\end{figure*}

We applied the K-S test to the observed and model
histograms using various combination of
($t_{\rm burst,min},t_{\rm burst,max}$) or ($t_{\rm stop,min},t_{\rm start}$),
putting constraints on these model parameters. A number of example
histograms are presented in Fig.~\ref{fig:cumhist}. The basic results are as 
follows:

\begin{itemize}
\item {\it All galaxies}: The observed cumulative histogram of colour
residuals has a clear continuous blue tail down to $\Delta(U-V)\sim-2$.
This allows for the existence of a fraction of young galaxies, and
models that continue to produce galaxies until $z=0.9$ are acceptable.
However, half of the galaxies are very red with $|\Delta(U-V)|<0.2$.
In order to make these red galaxies, 
$t_{\rm burst,max}$ should be larger than $\sim$ 4.5~Gyr in the single burst
model, or $t_{\rm start}$ should be larger than $\sim$ 3.5~Gyr in the constant
star formation model at 3 $\sigma$ level.
The former case corresponds to forming more than 40 per cent of the galaxies
beyond $z>2$, while the latter case corresponds to forming more than 40
percent of stars therein in total at $z>2$.
However note that, in any case, we do find a substantial fraction of galaxies
in which {\it all} the stars would have been formed at redshifts greater
than 2.
The best fitting model is
(single-burst, $t_{\rm burst,min}=0$~Gyr, $t_{\rm burst,max}=6$~Gyr).
In this case, $\sim$ 60 per cent of the galaxies are older than $z>2$.

\item {\it Cosmology}:
The $q_0=0.1$ cosmology is preferred because the position of the
ridge line matched the reddest colours expected from passive evolution. 
If we use a $q_0=0.5$ cosmology, the evolutionary correction makes
the Coma CMR too blue with respect to the HDF early-type
galaxies. Note, however, that the colours of
these galaxies are still consistent with the Coma CMR
within $\sim$ 1~$\sigma$ K-correction and photometric redshift errors.

\item {\it Bright galaxies}: 
In this case, we use a faint cut-off of $M_V-\log h<-19.8$
($q_0=0.1$), which roughly corresponds to 1 magnitude
fainter than the local $M_*$ (passively evolved out to $z = 0.9$).
The blue tail is strongly suppressed since most of the younger galaxies
are also faint.
Hence the fraction of old galaxies ($z>2$) is increased to more than
50 per cent. Their colour distribution is best fitted by
(single-burst, $t_{\rm burst,min}=1$-2~Gyr, $t_{\rm burst,max}=6$-5~Gyr) or
(constant-SF, $t_{\rm stop,min}=0$-1~Gyr, $t_{\rm start}=6$-5~Gyr).
In that case 70-80 per cent of galaxies or stars therein in total are older
than $z>2$.
It is also worth pointing out that galaxies bluer than $U-V=0.5$ will fade
by at least one magnitude {\it more} than the old passively evolving galaxies
before reaching the red envelope of the CMR.
Thus none of the blue early-type galaxies could be the progenitor of a
red object brighter than $M_V-\log h \sim -18.5$.

\item {\it No peculiar galaxies}:
F98's classification of the galaxies is mainly based on their surface
brightness profile. We checked the HST frames in order to visually confirm
the classification. Several galaxies were found to have morphological 
peculiarities (Table~\ref{tab:peculiar}). Most of these objects have
bluer colours. With these objects excluded, the colour distribution looks
quite similar to that of the bright galaxies, and the same arguments can be
applied.
\end{itemize}

\begin{table}
  \caption{Peculiar galaxies. Magnitudes and colours are given in the rest
 frame at $z=0.9$ same as in Fig.~\ref{fig:cmr_q01} ($q_0=0.1$).}
  \label{tab:peculiar}
  \begin{tabular}{lrrl}
  \hline
   FOCAS & $M_V$ & $U-V$ & Comments\\
   ID & $-5\log h$ & & \\
  \hline
  4-727-0 & $-$18.28 & $-$1.51 & Merger \\
  4-565-0 & $-$19.71 & 0.07 & Unclassified by F98 \\
  1-86-0  & $-$17.77 & $-$0.58 & Spiral structure\\
  4-307-0 & $-$20.77 & 1.22 & Faint and peculiar \\
  2-236-2 & $-$18.44 & 0.98 & Companion \\
  3-143-0 & $-$19.30 & 0.22 & Chain galaxy \\
  2-531-0 & $-$19.46 & 0.39 & Ring of \hii regions \\
  3-430-2 & $-$20.71 & 1.06 & Interaction \\
  3-405-0 & $-$18.03 & $-$1.21 & Spiral structure\\
  \hline
  \end{tabular}
  \medskip
\end{table}

In summary, the apparent tightness of the CMR is found to
be able to accommodate
continuous formation of early-type galaxies down to the observed epoch.
However, the redness of the galaxies is a considerable constraint,
and we found that at least 40 per cent of field early-type galaxies 
at $z = 0.9$ or stars therein in total (not in the individual galaxy)
should form beyond $z=2$ in order to create enough objects that are
sufficiently red.

\section{Discussion}

As we have described, the colour magnitude relation of early-type
galaxies in the HDF suggest a wide range of formation times, where
ongoing star formation is allowed in some galaxies right up to the 
epoch of observation.  However, the well-defined red ridge-line of
field early-type galaxies visible in Fig. \ref{fig:cmr_q01}
indicates that a substantial fraction of galaxies must have been
formed prior to $z > 2$.  

Our conclusions are consistent with the recent work on the HDF galaxies
by Abraham et al. (1998),
who concluded that 7 out of 11 $I_{814}$-band selected early-type
galaxies are coeval and old, and 4 galaxies have had recent star formation.
F98, from which we take our sample, claim 
that most of these galaxies have ages between 1.5 and 3.0~Gyr,
corresponding to $z_{\rm form}$ between 1 and 3 ($q_0=0.15$) based on the
rest-frame $B-J$ and $V-K$ colours. Although their upper limit on 
$z_{\rm form}$ is slightly lower, they do have a mixture of 
both young and old populations.
They also find that lower-mass systems preferentially form at lower redshifts,
the same tendency as we have found.

The substantial fraction of red galaxies raises two questions.
Firstly, is the presence of such red galaxies compatible with theories
of hierarchical galaxy formation? Although Governato et al. (1998) have shown
that early star formation happens in a fraction of objects, these
galaxies are likely to be subsequently bound into clusters.
Field early-type galaxies should have systematically younger ages
than their cluster counterparts (Kauffmann 1996). To
definitively rule out a particular model, however, requires the
selection process to be carefully taken into account. For instance,
the star formation histories of early-type galaxies in the local
field may differ substantially from early-type galaxies found in the
field at $z=0.9$. Furthermore, although our results suggest that
nearly 60 per cent of early-type galaxies 
were formed prior to $z=2$, the stars in these galaxies 
form only a small fraction
of the present day early-type population, and need not have been formed in
a single monolithic unit. This is an extensive topic that we will
investigate in a companion paper.

It is interesting to try to understand these results in the context of 
the evolution of the $K$-band luminosity function 
(Songaila et al. 1994; Cowie et al. 1996) and the space densities of red
galaxies in surveys such as the Canada-France Redshift Survey
(Lilly et al. 1995).
The decreasing space density of red galaxies (relative
to passively evolving models) have been shown to support the
hierarchical picture, indicating a late formation epoch for the majority
of field galaxies. 
Deep optical and NIR surveys find few objects red enough to be
consistent with the colours of a $z > 1$ passively evolving stellar
population, again supporting a late formation epoch for field galaxies
(Zepf 1997, Barger et al.\ 1998).  
However, optical-NIR colours are very sensitive to even small
amounts of star formation,  making these constraints relatively weak.
The evolution of the $K$-band luminosity function
is a particularly strong constraint (Kauffmann \& Charlot 1998b) since
it is robust to small amounts of residual star formation and has been
found to be consistent with the expectations from the hierarchical
clustering models.
In this context, our results, which
indicate a substantial old early-type population at $z \sim 0.9$, seem
to contradict these studies.

It is possible to reconcile these apparently contradictory observations
in a number of ways.  It is possible that the explicit selection of
{\it morphologically} early-type galaxies preferentially excludes
galaxies with ongoing star formation.  We would therefore be in a
situation where the field early-type galaxies at $z = 0.9$ are
merely a subset of the present-day field early-type galaxies, which
form from galaxies both with and without ongoing star formation at $z =
0.9$.  Alternatively, F98 suggest that the disappearance of 
red early-type galaxies may be due to the onset of an epoch of extreme dust
reddening at $z>1.5$, but such a picture seems ad hoc. Another 
solution may be to allow a degree of merging between the early-type
galaxies (eg., Menanteau et al. 1998). 
This would not contradict our conclusions concerning
the stellar populations, and, although the process would be expected to 
introduce additional scatter into the CMR, the observed relation
is sufficiently broad that this constraint is unimportant. 
Merging tends to flatten the CMR (BKT98), and in this respect it is
intriguing that the HDF relation is so flat. 
We note that another resolution to this discrepancy could be the
presence of a $z\sim1$ cluster in the HDF itself. Our conclusions would then
not be representative of the field (in the sense of representing an
average line of sight).
Although this seems unlikely, it can only be tested by extending the area
over which the properties of distant early-type field galaxies can be
studied. A first step in this direction will be the observation of the
Hubble Deep Field South (Williams et al. 1997).

\section{Conclusions}

Using the stellar population models and photometric redshift estimators
of Kodama et al.\ (1998a, 1998b), we constructed a
colour-magnitude diagram for morphologically selected HDF early-type
galaxies (as selected from F98) in the rest frame of the median
redshift of the sample $z = 0.9$.  The colour-magnitude diagram
constructed in this way has intrinsic scatter no smaller than the $z = 0.9$
cluster GHO 1603+4313.
Comparison of the
residuals from the passively evolved Coma CMR with simple models
indicate that a fraction of HDF early-types (primarily the low mass
galaxies) can form continuously to the epoch of observation.  However,
around half of our sample must form at $z > 2$ (although not
necessarily in a monolithic unit).  Reconciling this (high) fraction of
old field early-types at $z = 0.9$ with studies supporting late field
galaxy formation (as in HGF models) may be possible by taking into
account the morphological filtering of galaxies going into our CMR, or
by relatively late merging of smaller galaxies to produce our observed
early-type field sample.

\section*{Acknowledgements}

Most of this work was carried out during the Guillermo Haro Workshop '98 at
INAOE, Mexico. We thank all the organisers of this meeting for the invitation,
their kind hospitality, and the financial support to attend the meeting.
We are also grateful to the TMR Network Meeting on Galaxy Formation and
Evolution funded by the European Commission for providing us an opportunity
to develop this work. 
We would like to acknowledge helpful discussions with R. S. Ellis and
A. Arag\'on-Salamanca.
TK also thanks JSPS Postdoctoral Fellowships for Research Abroad for
financial support.
EFB would like to thank the Isle of Man Education Department for their
generous support.
This project made use of STARLINK computing facilities at Durham, Cambridge,
and INAOE.


\begin{thebibliography}{}

\bibitem[]{} Abraham, R. G., Ellis, R. S., Fabian, A. C., Tanvir, N. R.,
             \& Glazebrook, K., 1998, MNRAS, preprint
\bibitem[]{} Barger, A. J., Cowie, L. L., Trentham, N., Fulton, E., Hu,
             E. M., Songaila, A., \& Hall, D., 1998, AJ, in press, astro-ph/9809299
\bibitem[]{} Baugh, C. M., Cole, S., Frenk, C. S., \& Lacey, C. G., 1998,
             ApJ, 498, 504
\bibitem[]{} Bower, R. G., Lucey, J. R., \& Ellis, R. S., 1992, MNRAS, 254,
             601 (BLE92)
\bibitem[]{} Bower, R. G., Kodama, T., \& Terlevich, A, 1998, MNRAS, in press
             (BKT98)
\bibitem[]{} Butcher, H. \& Oemler, A., 1984, ApJ, 285, 426
\bibitem[]{} Cohen, J. G., Cowie, L. L., Hogg, D. W., Songaila, A.,
         Blandford, R., Hu, E. M., \& Shopbell, P., 1996, ApJ, 471, 5
\bibitem[]{} Cole, S., Arag\'on-Salamanca, A., Frenk, C. S., Navarro, J. F.,
             \& Zepf, S. E., 1994, MNRAS, 271, 781
\bibitem[]{} Cowie, L.L., 1998, unpublished catalogue of HDF redshifts,
        http://www.ifa.hawaii.edu/$^{\sim}$cowie/tts/tts.html
\bibitem[]{} Cowie, L.L., Songaila, A., Hu, E.M., \& Cohen, J.G., 1996,
             AJ, 112, 839
\bibitem[]{} Dickinson, M., et al., 1997, in preparation
\bibitem[]{} Ellis, R. S., Smail, I., Dressler, A., Couch, W. J., Oemler, A., 
        Butcher, H., \& Sharples, R. M., 1997, ApJ, 483, 582
\bibitem[]{} Fasano, G., Christiani, S., Arnouts, S., \& Filippi, M., 1998,
        AJ, in press
\bibitem[]{} Franceschini, A., Silva, L., Fasano, G., Granato, G. L.,
             Bressan, A., Arnouts, S., \& Danese, L., 1998, ApJ, in press (F98)
\bibitem[]{} Governato, F., Gardner, J. P., Stadel, J., Quinn, T., \& Lake, G.,
        1998, preprint, astro-ph/9710140
\bibitem[]{} Guzman, R., Lucey, J. R., Carter, D., \& Terlevich, R. J., 1992,
             MNRAS, 257, 187
\bibitem[]{} Kauffmann, G., White, S. D. M., \& Guiderdoni, B., 1993,
        MNRAS, 264, 201
\bibitem[]{} Kauffmann, G., 1996, MNRAS, 281, 487
\bibitem[]{} Kauffmann, G., \& Charlot, S., 1998a, MNRAS, 294, 705
\bibitem[]{} Kauffmann, G., \& Charlot, S., 1998b, MNRAS, 297, L23
\bibitem[]{} Kodama, T., \& Arimoto N., 1997, A\&A, 320, 41
\bibitem[]{} Kodama, T., Arimoto, N., Barger, A. J., \& Arag\'on-Salamanca, A.,
     1998a, A\&A, 334, 99
\bibitem[]{} Kodama, T., Bell, E. F, \& Bower, R. G., 1998b, MNRAS, submitted
\bibitem[]{} Kuntschner, H., \& Davies R. L., 1998, MNRAS, 295, 29
\bibitem[]{} Larson, R. B., Tinsley, B. M., \& Caldwell, C. N., 1980,
        ApJ, 237, 692
\bibitem[]{} Lilly, S. J., Tresse, L., Hammer, F., Grampton, D., \&
        Le Fevre, O, 1995, ApJ, 455, 108
\bibitem[]{} Menanteau, F., Ellis, R. S., Abraham, R. G., Barger, A. J.,
        \& Cowie, L. L., 1998, submitted.
\bibitem[]{} Stanford, S. A., Eisenhardt, P. R. M., \& Dickinson, M., 1998,
     ApJ, 492, 461
\bibitem[]{} Schmidt, M., 1959, ApJ, 129, 243
\bibitem[]{} Songaila, A., Cowie, L. L., Hu, E. M., \& Gardiner, J. P., 1994,
             ApJS, 94, 461
\bibitem[]{} Terlevich, A., 1998, PhD Thesis, University of Durham
\bibitem[]{} Visvanathan, N., \& Sandage, A., 1997, ApJ, 216, 214
\bibitem[]{} White, S. D. M., \& Frenk, C. S., 1991, ApJ., 379, 52
\bibitem[]{} Williams, R. E., et al., 1996, AJ, 112, 1335
\bibitem[]{} Williams, R. E., Baum, S. A., Blacker, B. S., et al., 
         1997, AAS, 191, 8508                   
\bibitem[]{} Worthey, G., 1994, ApJS, 95, 107
\bibitem[]{} Yee, H. K. C., 1998, astro-ph/9809347
\bibitem[]{} Zepf, S. E., 1997, Nature, 390, 377
\bibitem[]{} Zepf, S. E., Whitmore, B. C., \& Levison, H. F., 1991, ApJ, 383, 542

\end{thebibliography}
\end{document}